\documentstyle[preprint,aps,eqsecnum,epsf]{revtex}
\epsfverbosetrue

\ifx\epsffile\undefined
\def\insertfig#1#2{}
\def\insertfigsmall#1#2{}
\else
\def\insertfig#1#2{{\baselineskip=4pt
\centerline{\epsfxsize=\hsize\epsffile{#2}}{{
\centerline{#1}}}\medskip
}}
\def\insertfigsmall#1#2{{\baselineskip=4pt
\centerline{\epsfxsize=4in\epsffile{#2}}{{
\centerline{#1}}}\medskip
}}
\fi

\begin{document}

\def\dback{\raise1.5ex\hbox{$\leftharpoonup$}\mkern-16.5mu {\rm D}}
\def\dbacksl{\dback\hskip-0.6em /}
\def\dsl{{\rm D}\hskip-0.6em /}
\def\im{{\rm i}}
\def\matelement#1#2#3{\langle #1\vert #2\vert #3 \rangle}
\def\vdotvp{v\cdot v^\prime}
\def\slash#1{#1\hskip-0.7em /}
\def\vslash{{v\hskip-0.5 em /}}
\def\kslash{{k\hskip-0.5 em /}}
\def\lqcd{\Lambda_{\rm QCD}}
\def\msbar{{\overline{\rm MS}}}
\def\OMIT#1{}

{\tighten
\preprint{\vbox{\hbox{UTPT 94-21}
\hbox{UCSD/PTH 94--12}
\hbox{CMU-HEP 94--21}
\hbox{hep-ph/9407407}
}}
\title{Renormalons in Effective Field Theories}
\author{Michael~Luke}
\address{Department of Physics,
University of Toronto, Toronto Canada M5S 1A7}
\author{Aneesh V. Manohar}
\address{Department of Physics, University of California at San Diego,
La Jolla, CA 92093}
\author{Martin J. Savage}
\address{Department of Physics, Carnegie Mellon University, Pittsburgh, PA
15213}
\bigskip
\date{July 1994}
\maketitle
\begin{abstract}
We investigate the high-order behavior of perturbative matching conditions
in effective field theories.  These series are typically badly divergent,
and are not Borel summable due to infrared and ultraviolet renormalons
which introduce ambiguities in defining the sum of the series.  We argue
that, when treated consistently, there is no physical significance to these
ambiguities.  Although nonperturbative matrix elements and matching
conditions are in general ambiguous, the ambiguity in any physical
observable is always higher order in $1/M$ than the theory has been
defined.  We discuss the implications for the recently noticed infrared
renormalon in the pole mass of a heavy quark.  We show that a ratio of form
factors in exclusive $\Lambda_b$ decays (which is related to the pole mass)
is free from renormalon ambiguities regardless of the mass used as the
expansion parameter of HQET.  The renormalon ambiguities also cancel in
inclusive heavy hadron decays.  Finally, we demonstrate the cancellation of
renormalons in a four-Fermi effective theory obtained by integrating out a
heavy colored scalar.
\end{abstract}

\pacs{13.20.He, 12.38.Bx, 13.20.Fc, 13.30.Ce}
}

\section{Introduction}
In physical problems involving several distinct scales, it is often
convenient to describe the physics using an effective field theory.
Typically, one is interested in physics at energies much less than the mass
of a heavy particle, in which case the physics is most easily described
by an effective Lagrangian in which virtual heavy particle exchange is
accounted for through a series of non-renormalizable operators.  The
coefficients of these operators are perturbatively calculable as a power
series in $\alpha_s(M)$, where $M$ is the mass of the heavy particle which
is integrated out.  However, since the resulting perturbation series is
only asymptotic, and since it is well known that perturbative QCD is not
Borel-summable, the perturbatively calculated coefficient functions are at
some level ambiguous.  One source of this ambiguity is infrared
renormalons\cite{thooft,parisi,david,mueller}, which arise in QCD from
graphs of the form shown in Fig.~1.  They contribute to the factorial
growth of the coefficients of the perturbation series, and introduce
uncertainties in the sum of the perturbation series proportional to
powers of $\Lambda_{QCD}/M$.

There has been much recent discussion in the literature on the effects of
infrared renormalons in the context of one particular effective field
theory, the heavy quark effective theory (HQET)
\cite{bigietal,benekebraun,bbz}.  In particular, the use of the pole
mass as an expansion parameter in HQET has been criticized, because it
has been shown to suffer from an ambiguity which prevents its definition
to an accuracy better than ${\cal O}(\Lambda_{\rm QCD})$ \cite
{bigietal,benekebraun}, and the authors of Ref.~\cite{bigietal} advocate
formulating HQET in terms of a short-distance mass $m(\mu)$ which does
not suffer from a renormalon  ambiguity at ${\cal O}(\Lambda_{\rm
QCD})$. The results of Ref.~\cite{bigietal,benekebraun} clarify the
formal status of the pole mass (and of nonperturbative matrix elements
in an effective theory in general).

Given this ambiguity, it is important to show that physical predictions
of HQET, which are usually expressed using the pole mass as an expansion
parameter, are unambiguous.   In this paper, we investigate the effects
of renormalons on matching conditions in effective field theories.  We
argue that, while perturbatively calculated coefficient functions suffer
from renormalons, any ambiguity in a physical observable is always
higher order in $1/M$ than the theory has been defined and is
consequently irrelevant.  Therefore, as long as one works consistently,
it does not matter that unobservable parameters such as the heavy quark
mass or the matrix elements of higher dimension operators are not
unambiguously defined; relations between physical quantities are
unambiguous.  We also argue that, while formulating HQET in terms of
some short-distance mass $m(\mu)$ is certainly possible,  use of an
expansion parameter  other than the pole mass (or some mass which
differs from $m_{pole}$ by ${\cal O}(\Lambda_{QCD})$) is both
inconvenient and uneccessary.

The paper is arranged as follows: In Section~2, we review the physics of
infrared renormalons, and discuss the basic method of calculation that will
be used in this paper.  In Section~3 we discuss HQET and the renormalon
ambiguity in the quark pole mass.  We show explicitly that there are no
renormalons in the ratio of form factors for $\Lambda_b\rightarrow
\Lambda_c$ semileptonic decay from which the perturbatively defined pole
mass may be extracted, and we comment on the renormalon cancellation for
inclusive semileptonic $B$ meson decay.  In Section~4, we discuss
renormalon ambiguities in a four-Fermi effective theory.  We show that this
effective theory has new renormalon ambiguities not present in the full
theory which cancel corresponding ambiguities in matching conditions,
and also show that the cancellation of renormalon ambiguities is not specific
to HQET, but occurs in other effective field theories.  This section can
be omitted by readers only interested in HQET. The conclusions are
presented in Section~5.

\section{Renormalons}

QCD perturbation theory is used to express some quantity $f$ as a power
series in $\alpha_s$,
\begin{equation}\label{fseries}
f(\alpha_s) = f(0)+ \sum_{n=0}^\infty f_n\ \alpha_s^{n+1}.
\end{equation}
Typically, this perturbation series for $f$ is only asymptotically
convergent. The convergence can be improved by defining the Borel transform
of $f$,
\begin{equation}\label{borel}
  B[f](t)=f(0)\ \delta(t)+\sum_{n=0}^\infty {f_n\over n!}\ t^n,
\end{equation}
which is more convergent than the original expansion eq.~(\ref{fseries}). The
original expression $f(\alpha_s)$ can be recovered from the Borel transform
$B[f](t)$ by the inverse Borel transform
\begin{equation}\label{inverse}
  f(\alpha_s)=\int_0^\infty dt\ e^{-t/\alpha_s}\ B[f](t).
\end{equation}
If the integral in eq.~(\ref{inverse}) exists, the perturbation series
$f(\alpha_s)$ is Borel-summable, and is unambiguously defined.  However, if
there are singularities in $B[f](t)$ along the path of integration, the
function $f$ is ambiguous.  The inverse Borel transform must be defined by
deforming the contour of integration away from the singularity, and the
inverse Borel transform in general depends on the deformation used.

One source of singularities in $B[f]$ in QCD is infrared renormalons
\cite{thooft,parisi,david,mueller}.  These arise from graphs of the form in
Fig.~1.  Physically, these graphs correspond to the running of $\alpha_s$,
and infrared renormalons are ambiguities in perturbation theory arising
from the fact that the gluon coupling gets strong for soft gluons in the
one-loop diagram in Fig.~1.  The infrared renormalons produce a factorial
growth in the coefficients $f_n$, which gives rise to poles in the Borel
transform $B[f]$.  The renormalon ambiguities have a power law dependence
on the momentum transfer $Q^2$.  For example, a simple pole at $t=t_0$
in $B[f]$ introduces an ambiguity in $f$ depending on whether the
integration contour is deformed to pass above or below the renormalon
pole.  The difference between the two choices is proportional to
\begin{eqnarray}\label{rensing}
  \delta f &\sim& \oint_C e^{-t_0/\alpha_s(Q)}\ B[f](t)\\
   &\sim& \left(\lqcd\over Q\right)^{2 (-b_0) t}, \nonumber
\end{eqnarray}
where $b_0=-(11-2n_f/3)/4\pi$ is the leading term in the QCD $\beta$-function
\begin{equation}\label{betafn}
\mu^2 {d\alpha_s\over d\mu^2} = b_0\alpha_s^2 + {\cal O}\left(
\alpha_s^3 \right),
\end{equation}
which governs the high energy behavior of the QCD coupling constant
\begin{equation}\label{largeqalpha}
\alpha_s(Q) = {1\over (-b_0) \ln\left(Q^2/\lqcd^2\right)},
\end{equation}
and the contour $C$ encloses $t_0$.
It is useful to write the Borel transform $B[f](t)$ in terms of the
variable $u=-b_0t$.\footnote{This is the definition used in
\cite{benekebraun}, and is the negative of the definition used in
\cite{beneke}.} The form of the renormalon singularity in
eq.~(\ref{rensing}) then implies that a renormalon at $u_0$ produces an
ambiguity in $f$ that is of order $\left(\lqcd/Q\right)^{2u_0}$.  This
ambiguity is canceled by a corresponding ambiguity in non-perturbative
effects such as in the matrix elements of higher dimension operators.
The sum of the perturbation series plus non-perturbative corrections is
expected to be well defined.

\subsection*{The Calculational Method}

Clearly, one cannot sum the entire QCD perturbation series to determine the
renormalon singularities.  Typically, one sums bubble chains of the form
given in Fig.~1 \cite{thooft,lautrup}.  Beneke \cite{beneke} considered a
limiting case of QCD in which the bubble chain sum is the leading
contribution to the renormalon.  Take QCD with $N_f$ flavors in the limit
$N_f \rightarrow \infty$ with $a=N_f\alpha_s$ held fixed.  Feynman diagrams
are computed to leading order in $\alpha_s$, but to all orders in $a$.
Terms in the bubble sum of Fig.~1 with any number of bubbles are equally
important in this limit, since each additional fermion loop contributes a
factor $\alpha_s N_f$, which is not small.  QCD is not an asymptotically
free theory in the $N_f\rightarrow \infty$ limit, so the procedure used by
Beneke is to write the Borel transform as a function of $u=-b_0 t$ but
still study renormalons for positive $u$.  The singularities in $u$ are
taken to be the renormalons for asymptotically free QCD.  This procedure is
a formal way of doing the bubble chain sum, while neglecting other
diagrams.

The Borel transform of the sum of Feynman graphs containing a single bubble
chain can be readily obtained by performing the Borel transform before doing
the loop integral
\cite{beneke,benekebraun}. The bubble chain sum is
\begin{equation}\label{chainsum}
G(\alpha_s,k) = \sum_{n=0}^\infty \left({k_\mu k_\nu\over k^2} -
g_{\mu\nu}\right)
(b_0 \alpha_s N_f)^n  \left( \ln(-k^2/\mu^2) + C\right)^n
\end{equation}
where $k$ is the momentum flowing through the gauge boson propagator, $C$
is a constant that depends on the particular subtractions scheme, and
$b_0=1/6\pi$ is the contribution of a single fermion to the
$\beta$-function.  In the ${\overline{\rm MS}}$ scheme, $C=-5/3$.  The Borel
transform
of eq.~(\ref{chainsum}) with respect to $\alpha_s N_f$ is ($u=-b_0t$)
\begin{eqnarray}\label{borelchain}
B[G](u,k) &=& {1\over \alpha_s N_f} \sum_{n=0}^\infty {1\over k^2}\left({k_\mu
k_\nu\over k^2} - g_{\mu\nu}\right) {(-u)^n\over n!}  \left( \ln(-k^2/\mu^2) +
C\right)^n
\nonumber \\ &=&
{1\over \alpha_s N_f} {1\over k^2} \left({k_\mu k_\nu\over k^2} -
g_{\mu\nu}\right)
\exp\left[-u \ln(-k^2 e^C/\mu^2) \right] \\ &=& {1\over \alpha_s N_f} \left(
{\mu^2\over e^C} \right)^u { 1\over \left(-k^2\right)^{2+u}} \left({k_\mu
k_\nu}
- k^2
g_{\mu\nu}\right).
\nonumber
\end{eqnarray}
The Borel transformed loop graphs can be computed by using the propagator in
eq.~(\ref{borelchain}) instead of the usual gauge boson propagator
\begin{equation}
\left({k_\mu k_\nu} - k^2 g_{\mu\nu}\right){1\over \left(k^2\right)^2}.
\end{equation}

\section{Renormalons in The Heavy Quark Effective Theory}

\subsection{Matching Conditions}
An effective field theory Lagrangian is an expansion in an operator series
in inverse powers of some mass scale $M$.  By construction, the effective
field theory has the same infrared physics as the full theory.  However,
because the ultraviolet physics (above the scale at which the theories are
matched) differs dramatically in the two theories, the coefficients of
operators in the effective theory must be modified at each order in
$\alpha_s(M)$ to ensure that physical predictions are the same in the two
theories.

Since the two theories coincide in the infrared, these matching conditions
depend in general only on ultraviolet physics and should be independent of
any infrared physics, including infrared renormalons.  However, in a
mass-independent renormalization scheme such as dimensional regularization
with ${\overline {\rm MS}}$, such a sharp separation of scales cannot be
achieved.  It is easy to see how infrared renormalons creep into matching
conditions.  Consider the familiar case of integrating out a $W$ boson and
matching onto a four-Fermi interaction (we will discuss a variant of this
example in detail in Section~4).  The matching conditions at one loop
involve subtracting one-loop scattering amplitudes calculated in the full
and effective theories, as indicated in Fig.~2.  For simplicity, neglect
all external momenta and particle masses, and consider the region of loop
integration when the gluon is soft.  When $k=0$, the two theories are
identical and the graphs in the two theories are identical.  This is the
well-known statement that infrared divergences cancel in matching
conditions.  However, for finite (but small) $k$, the two theories differ
at ${\cal O}(k^2/M_W^2)$ when one retains only the lowest dimension operators
in the effective theory.  Therefore, the matching conditions are sensitive to
soft gluons at this order, and it is not surprising that (as we shall show)
the resulting perturbation series is not Borel-summable and has renormalon
ambiguities starting at ${\cal O}(\Lambda_{\rm QCD}^2/M_W^2)$.

However, this ambiguity is completely spurious, and does {\it not} mean
that the effective field theory is not well defined.  Since the theory has
only been defined to a fixed order, an ambiguity at higher order in $1/M_W$
is irrelevant.  The renormalon ambiguity corresponded to the fact that the
two theories differed in the infrared at ${\cal O}(k^2/M_W^2)$.  When operators
suppressed by an additional power of $1/M_W^2$ in the effective theory
are consistently taken into account, the two theories will coincide in
the infrared up to ${\cal O}(k^4/M_W^4)$, and any ambiguity is then
pushed up to ${\cal O}(\Lambda_{QCD}^4/M_W^4)$.  Consistently including
$1/M_W^4$ suppressed operators pushes the renormalon to ${\cal
O}(\Lambda_{QCD}^6/M_W^6)$, and so on.  In general, a renormalon at
$u=u_0$ in the coefficient function of a dimension $D$ operator is
canceled exactly by a corresponding ambiguity in matrix elements of
operators of dimension $D-4+2u_0$, so that physical quantities are
unambiguous.  This cancellation is a generic feature of all effective
field theories, and also occurs in HQET.

\subsection{HQET and the Quark Mass}

The HQET Lagrangian has an expansion in
inverse powers of the heavy quark mass,
\begin{eqnarray}\label{hqetLagrangian}
  {\cal L}&=&{\cal L}_0+{1\over 2m_0} {\cal L}_1 + {1\over (2m_0)^2}{\cal
L}_2 +
\ldots +
  {\cal L}_{\rm light},\nonumber \\
  \noalign{\smallskip}
  {\cal L}_0&=&\bar h_v\, (i{\rm D}\cdot v)\, h_v-\delta m\ \bar h_v h_v ,\\
  {\cal L}_1&=& \ldots\ . \nonumber
\end{eqnarray}
Here ${\cal L}_{\rm light}$ is the QCD Lagrangian for the light quarks and
gluons, $h_v$ is the heavy quark field, and ${\cal L}_k$ are terms in the
effective Lagrangian for the heavy quark that are of order $1/m_0^k$.
There are two mass parameters for the heavy quark in
eq.~(\ref{hqetLagrangian}), the expansion parameter of HQET $m_0$, and the
residual mass term $\delta m$.  The two parameters are not independent; one
can make the redefinition $m_0\rightarrow m_0 + \Delta m$, $\delta m
\rightarrow \delta m - \Delta m$.  A particularly convenient choice is to
adjust $m_0$ so that the residual mass term $\delta m$ vanishes.  Most HQET
calculations have been done with this choice of $m_0$, but it is easy to
show that the same results are obtained with a different choice of $m_0$
\cite{fnl}.  The HQET mass when $\delta m=0$ is often referred to in the
literature as the pole mass, and we will follow this practice here.

Like all effective Lagrangians, the HQET Lagrangian is non-renormalizable,
so a specific regularization prescription must be included as part of the
definition of the effective theory.  An effective field theory is used to
compute physical quantities in a systematic expansion in a small parameter,
and the effective Lagrangian is expanded in this small parameter.  The
expansion parameter of the HQET is $\lqcd/m_0$.  One can then use ``power
counting'' to determine what terms in the effective theory are relevant to
a given order in the $1/m_0$ expansion.  For example, to second order in
$1/m_0$, one needs to study processes to first order in ${\cal L}_2$, and to
second order in ${\cal L}_1$.  It is important that the renormalization
procedure preserve the the power counting for the effective field theory to
make sense.  In order to preserve power counting, we must choose a
mass-independent subtraction scheme; in our case, we choose to use
dimensional regularization and ${\overline{\rm MS}}$.  A mass-dependent
subtraction scheme (such as a momentum space cutoff) mixes different
orders in the $1/m_0$ expansion.  Thus to compute a quantity to first
order in $1/m_0$, one would have to retain the effective Lagrangian to
all orders in $1/m_0$, which is not particularly useful.  Hence, HQET is
defined using a mass-independent subtraction scheme, and nonperturbative
matrix elements must be interpreted in this scheme.

It has recently been shown \cite{bigietal,benekebraun} that there is a
renormalon in the relation between the renormalized mass at short distances
(such as the ${\overline{\rm MS}}$ mass $\overline m$) and the pole mass of the
heavy quark at $u=1/2$, which produces an ambiguity in the relation between the
pole mass and the ${\overline{\rm MS}}$ mass of order $\lqcd$.  This implies
that there is an ambiguity in the residual mass term $\delta m$ of order
$\lqcd$ due to renormalon effects \cite{bigietal,benekebraun}.

The quark mass in HQET and the ${\overline{\rm MS}}$ mass at
short-distances are parameters in the Lagrangian that must be determined
from experiment.  Any scheme can be used to compute physical processes,
though one scheme might be more advantageous for a particular
computation.  The ${\overline{\rm MS}}$ mass at short distances is
useful in computing high-energy processes.  However, there is no
advantage to using the ``short distance'' mass (such as the running
${\overline{\rm MS}}$ mass) in HQET, as advocated by \cite{bigietal}. In
fact, from the point of view of HQET, this is extremely inconvenient.  The
effective Lagrangian eq.~(\ref{hqetLagrangian}) is an expansion in inverse
powers of $m_0$.  Power counting in $1/m_0$ in the effective theory is only
valid if $\delta m$ is of order one (or smaller) in $m_0$, i.e.  {\it only}
if $\delta m$ remains finite in the infinite mass limit $m_0 \rightarrow
\infty$.  When $m_0$ is chosen to be the ${\overline{\rm MS}}$ mass the
residual mass term $\delta m$ is of order $m_0$ (up to logarithms),
which spoils the $1/m_0$ power counting of HQET, mixes the $\alpha_s$
and $1/m_0$ expansions, and breaks the heavy flavor symmetry.  For
example, using $m_0$ to be the ${\overline{\rm MS}}$ mass at
$\mu=m_0$, one finds at one loop that
\begin{equation}
      \delta m ={4\over 3\pi} \alpha_s m_0.
\end{equation}
In $b\rightarrow c$ decays, including this residual mass term in the heavy
$c$-quark Lagrangian, causes $1/m_c$ operators such as $\bar h_c
(-\im\dbacksl) \Gamma h_b/m_c$ to produce effects that are of the same
order in $1/m_c$ as lower dimension operators of the form $\bar h_c
\Gamma h_b$.  While physical quantities calculated in this way must be
the same as those calculated using the pole mass, it unnecessarily
complicates calculations to use a definition for $m_0$ that leaves a
residual mass term that is not finite in the $m_0 \rightarrow \infty$
limit.  Better choices of the expansion parameter of HQET are the heavy
meson mass (with $\delta m$ of order $\lqcd$), and the pole mass (with
$\delta m=0$).

The ${\overline{\rm MS}}$ mass at short-distances can be determined (in
principle) from experiment without any renormalon ambiguities
proportional to $\Lambda_{\rm QCD}/m_Q$ (i.e. at $u=1/2$).  As an
example, consider the ${\overline{\rm MS}}$ mass of the $b$-quark at
the GUT scale in an $SU(5)$ unified field theory.  The $b$-quark mass at
the GUT scale is proportional to the $b$-quark Yukawa coupling at the GUT
scale, which in turn is equal to the $\tau$-lepton Yukawa coupling at the
same scale\footnote{There are corrections to this relation from matching
conditions at the GUT scale, which will have renormalon ambiguities
proportional to powers of $\Lambda_{\rm QCD}/m_{\rm GUT}$.}.  There are no
QCD renormalons at $u=1/2$ in the relation between the $\tau$-lepton mass
at short-distances, and the pole mass of the $\tau$ (neglecting QED
effects).  Thus one could determine the $b$-quark mass at short distances
by measuring the $\tau$-lepton mass, without any renormalon ambiguities at
$u=1/2$.

The ${\overline{\rm MS}}$ quark mass can be related to other definitions of the
quark mass using QCD perturbation theory. The connection between the Borel
transformed pole mass and a short-distance mass (such as the ${\overline{\rm
MS}}$ mass) has been worked out in ref.~\cite{benekebraun}. The relation
between the two (for the $c$-quark) is
\begin{equation}\label{borelmass}
  B[m^{\rm pole}_c](u)= m_c\delta(u) + {m_c\over 3\pi N_f}\left[\left(
  \mu^2\over m_c^2\right)^u e^{-uC} 6(1-u){\Gamma(u)\Gamma(1-2u)\over
  \Gamma(3-u)}-{3\over u}+R_{\Sigma_1}(u)\right]
\end{equation}
where $m_c$ is the renormalized ({\it not} the pole) mass at short distances,
such as the ${\overline{\rm MS}}$ mass, $\mu$ is the renormalization scale, and
the constant $C$ and the function $R_{ \Sigma_1 } (u)$ depend on the
renormalization scheme. Eq.~(\ref{borelmass}) has a renormalon
singularity at $u=1/2$ which is the leading infrared  renormalon in the
pole mass. Writing $u=1/2+\Delta u$, we have
\begin{equation}\label{mpole}
  B[m_c^{\rm pole}](u=1/2+\Delta u)=-{2\mu e^{-C/2}\over 3\pi N_f  m_c\, \Delta
u}+\ldots
\end{equation}
where the ellipsis denotes terms regular at $\Delta u=0$. In the next
two sections, we will only work to ${\cal O}\left(1/m_0\right)$, so
poles to the right of $u=1/2$, which are related to ambiguities at
higher order in $1/m_0$, are irrelevant at this stage.

Although $m_c^{\rm pole} $ is formally ambiguous at ${\cal O}(1/m_c)$, we
will argue in  this paper that physical quantities which depend on $m_c^{\rm
pole}$ are unambiguously predicted in HQET.  We demonstrate this
explicitly for a ratio of form factors in $\Lambda_b$ semileptonic
decay.  We then comment on the cancellation of  renormalon ambiguities
in the expression for the inclusive semileptonic width of the $B$
meson. Both results will make use of eq.~(\ref{mpole}) and its analog
for the b-quark.

\subsection{$\bar\Lambda$ from Exclusive Decays}

The matrix element of the vector current for the semileptonic decay $\Lambda_b
\rightarrow \Lambda_c e^-\nu_e$ decay is
parameterized by the three decay form factors
\begin{equation}\label{formfactors}
  \matelement{\Lambda_c(v^\prime)}{\bar c\gamma^\mu b}{\Lambda_b(v)}=
  \bar u(v^\prime)\left[ F_1(\vdotvp)\gamma^\mu+F_2(\vdotvp) v^\mu
  +F_3(\vdotvp)v^{\prime \mu}\right] u(v).
\end{equation}
In the limit $m_b,\ m_c\rightarrow\infty$, and at lowest order in $\alpha_s$,
the form-factors $F_2$ and $F_3$ vanish. We will consider $\alpha_s$ and
$1/m_c$ corrections, but work in the $m_b=\infty$ limit. Consider the ratio
$r_F=F_2/F_1$, which vanishes at lowest order in $\alpha_s$ and $1/m_c$. The
corrections to $r_F$ can be written in the form \cite{ggw}
\begin{equation}\label{f2overf1}
  r_F(\alpha_s,\vdotvp)\equiv {F_2(\vdotvp)\over
  F_1(\vdotvp)}={\bar\Lambda\over m_c}{1\over(1+\vdotvp)}+
  f_r(\alpha_s,\vdotvp)
\end{equation}
where the function $f_v(\alpha_s,\vdotvp)$ is a perturbatively calculable
matching condition from the theory above $\mu=m_c$ to the effective theory
below $\mu=m_c$, and the $\bar\Lambda$ term arises from $1/m_c$ suppressed
operators in HQET. At one loop, \cite{fggw}
\begin{equation}
  f_r(\alpha_s,\vdotvp)=-{2\alpha_s\over 3\pi}
  {1\over\sqrt{(\vdotvp)^2-1}}\ln\left(
  \vdotvp+\sqrt{(\vdotvp)^2-1}\right).
\end{equation}

The ratio $r_F=F_2/F_1$ is an experimentally measured quantity, and does
not have a renormalon ambiguity.  The standard form for $r_F$ in
eq.~(\ref{f2overf1}) is obtained by using HQET with the pole mass as the
expansion parameter.  The HQET parameter $\bar\Lambda$ is the meson mass in
the effective theory, i.e.  it is the meson mass $m_D$ minus the pole mass
of the c-quark.  The pole mass has the leading renormalon ambiguity
\cite{benekebraun,bigietal} at $u=1/2$ given in eq.~(\ref{mpole}), which
produces an ambiguity in the $1/m_c$ contribution to $F_2/F_1$ given by the
first term in eq.~(\ref{f2overf1}).  There must therefore also be a
renormalon at $u=1/2$ in the radiative correction to $F_2/F_1$ given by the
second term in eq.~(\ref{f2overf1}).  It is straightforward to show, using the
techniques of Section~2, that this is indeed the case.

The Borel transformed series $B[f_r](u,\vdotvp)$ in the
$1/N_f$ expansion is easily calculated from the graph in Fig.~3 using the
Borel transformed propagator in eq.~(\ref{borelchain}).  The Borel
transform of the Feynman diagram is
\begin{equation}\label{diagram}
  B[graph]={1\over \alpha_s N_f}{4\over 3} g^2 \left({\mu^2\over e^C}\right)^u
  \int{{d^4k\over(2\pi)^4}\,
  {\gamma^\nu\left(m_c \vslash^\prime+\kslash+m_c\right)\gamma^\alpha v^\mu
  (k_\mu k_\nu-k^2 g_{\mu\nu})\over
  (k^2+2m_c k\cdot v^\prime)(-k^2)^{2+u} k\cdot v}}.
\end{equation}
The radiative correction to $F_2$ (which determines $f_r$) is obtained from
the terms in eq.~(\ref{diagram}) which are proportional to $v^\alpha$.
Combining denominators using the identities
\begin{eqnarray}
{1\over (k^2+2m_c k\cdot v^\prime)(k^2)^{2+u}} &=& (2+u) \int_0^1 {dx
\left(1-x\right)^{1+u} \over \left[ k^2 + 2m_c x k\cdot v^\prime\right]^{3+u}
},
\nonumber \\
{1\over \left(k^2+2m_c  x k\cdot v^\prime\right)^{3+u} k\cdot v} &=& 2 (3+u)
\int_0^\infty {d\lambda \over \left[ k^2 + 2m_c x k\cdot v^\prime +
2 \lambda k
\cdot v \right]^{4+u}}, \nonumber
\end{eqnarray}
extracting the terms proportional to $v^\mu$ and performing the momentum
integral, we obtain
\begin{equation}
B[f_r](u,\vdotvp)={4(u-2)\over 3 \pi N_f(1+u)} \left({\mu^2\over e^C}\right)^u
m_c
\int_0^\infty d\lambda \int_0^1 dx {\left(1-x\right)^{1+u} x \over
\left[ \lambda^2 + 2 \lambda m_c x  v \cdot v^\prime+ m_c^2 x^2
\right]^{1+u}}.
\end{equation}
Rescaling $\lambda \rightarrow x m_c \lambda$ and performing the
$x$ integral gives
\begin{equation}\label{intresult}
B[f_r](u,\vdotvp)={4\over 3 \pi N_f} \left({\mu^2\over m_c^2 e^C}\right)^u
{(u-2)\Gamma(1-2u)\Gamma(1+u)\over\Gamma(3-u)}\int_0^\infty d\lambda
{1\over
\left[ \lambda^2 + 2 \lambda  v \cdot v^\prime +1 \right]^{1+u}}.
\end{equation}
This expression has a pole at $u=1/2$. Expanding in  $\Delta u =u-1/2$ gives
\begin{eqnarray}\label{irrenormalon}
  B[f_r](u=1/2+\Delta u,\vdotvp)&=&{2\mu \over 3\pi m_c e^{C/2}}
\ {1\over \Delta u}\
\int_0^\infty
d\lambda {1\over
\left[ \lambda^2 + 2 \lambda  v \cdot v^\prime +1 \right]^{3/2}} +
...\nonumber
\\
&=&{2\mu \over 3\pi m_c e^{C/2}} {1\over \Delta u}{1\over 1+\vdotvp}
\end{eqnarray}
where the ellipsis denotes terms that are regular at $u=1/2$.

The Borel singularity in eq.~(\ref{irrenormalon}) cancels the singularity
in the first term of eq.~(\ref{f2overf1}) at all values of $v\cdot
v^\prime$, so that the  form factor ratio $r_F(\alpha_s, v \cdot
v^\prime) = F_2(v \cdot v^\prime)/F_1(v \cdot v^\prime)$ has no
renormalon ambiguities.  Therefore the standard HQET computation of the
$1/m_c$ correction to $F_2/F_1$ using the pole mass and the standard
definition of $\bar\Lambda$ gives an unambiguous physical prediction for
the ratio of form factors.

\subsection{Inclusive Decays}

A similar situation occurs for inclusive $B$ decays, which have been
the subject of much recent interest \cite{cgg,MW,Bigietc}.  The
inclusive $B\rightarrow X_q e\nu$ (where $q=u$ or $c$) decay rate is
related to the imaginary part of the forward scattering amplitude,
\begin{equation}\label{twocurrents}
  \Gamma(B\rightarrow X_q e\nu)\sim{\rm
Im}\ \matelement{B}{T(J^{\mu\dagger},J^\nu)}{B},
\end{equation}
where $J^\mu=\bar c\gamma^\mu(1-\gamma_5) b$.
In this case the expression for the total rate as an expansion in powers
of $1/m_Q$ is not the result of matching onto an effective theory, but
instead is the result of
performing an operator product expansion on the time ordered product
of the two currents in Eq.~(\ref{twocurrents}). The final expression is
\begin{eqnarray}\label{totalwidth}
  \Gamma(B\rightarrow X_q e\nu)&=&{G_f^2\, \vert V_{bq} \vert^2\,
  m_b^5\over 192
  \pi^3}\Bigl[ f_0(\alpha_s)\ \matelement{B(v)}{\ \bar h_b
  h_b\ }{B(v)}\nonumber\\
  &+&{5\over m_b} f_1(\alpha_s)\ \matelement{B(v)}{\ \bar h_b\, \im \left({\rm
  D}\cdot v\right) h_b\ }{B(v)}+{\cal
O}(m_q^2/m_b^2,\Lambda_{QCD}^2/m_b^2)\Bigr],
\end{eqnarray}
where $f_0=1$ and $f_1=1$ to lowest order in $\alpha_s$.
Eq.~(\ref{totalwidth}) is true with an arbitrary residual mass
term in the HQET Lagrangian, and we have not yet applied the equations of
motion to the operator $\bar h_b\, \im \left({\rm D}\cdot v\right) h_b$.
The total decay rate $\Gamma$ is an observable, and does not have a
renormalon ambiguity.  It was shown in Refs.~\cite{bigietal,bbz} that
the  total decay rate is unambiguous.  It is important to note that this
result  does not require the use of a ``short-distance" mass in
Eq.~(\ref  {totalwidth}).  One is free to choose some definite (but arbitrary)
prescription for integrating around the pole at $u=1/2$ in
Eq.~(\ref{borelmass}).  The mass $m_b$ in the leading term of
Eq.~(\ref{totalwidth}) is then well-defined, but there is an ambiguity
at ${\cal O}(\Lambda_{\rm QCD})$ in the residual mass term $\delta m$
arising from the renormalon at $u=1/2$ in the pole mass. By the
equations of motion,
\begin{equation}
  \im (v\cdot{\rm D})\ h_b=\delta m\ h_b+{\cal O}(1/m_b),
\end{equation}
so the ambiguity at ${\cal O}(\Lambda_{\rm QCD})$ in the matrix element
$\matelement{B(v)}{\ \bar h_b\, \im {\rm D}\cdot v h_b\ }{B(v)}$ is the
same as that in the pole mass, eq.~(\ref{mpole}). From \cite{bbz}, the
Borel transformed series $B[f_0(u)]$ for the $\alpha_s$ corrections to the
leading term has a pole at $u=1/2$,
\begin{equation}\label{increnormalon}
  B[f_0](u=1/2+\Delta u)={10\mu e^{-C/2}\over 3\pi N_f m_b \Delta
u}+\ldots
\end{equation}
 Comparing eq.~(\ref{mpole})
with eq.~(\ref{increnormalon}) and eq.~(\ref{totalwidth}), we see explicitly
that the ambiguity in the matrix element of $\im D\cdot v$ cancels that
in $f_0$ to give an unambiguous prediction for the total width $\Gamma$.

Thus in this O.P.E. the cancelation of renormalon ambiguities occurs in
the same manner as in the construction of HQET:  the ambiguity in the
matrix element of a higher dimension operator cancels that in the
perturbation series for the coefficient of the leading operator.

\section{Four-Fermi Theory}

As we argued in Section~3, the cancellation of renormalon ambiguities
between matrix elements and matching conditions is a general feature of an
effective field theory.  In this section, we illustrate this in a more
familiar effective field theory, four-Fermi theory in which a heavy colored
scalar is integrated out. (We choose the scalar theory as our example
because it has Feynman graphs which are slightly easier to compute than in
the four-Fermi theory of weak interactions.) The theory is QCD with $N_f$
light flavors in the $N_f \rightarrow \infty$ limit, with $\alpha_s N_f$
fixed.  Two of the $N_f$ flavors (called $d$ and $s$) couple via a color
triplet scalar of mass $M\gg\lqcd$ to color singlet particles (called $e$
and $\tau$) according to the interaction Lagrangian
\begin{equation}\label{slag}
  {\cal L}= \lambda\left(\bar s \tau + \bar d e\right) \phi + {\rm h.c.}\ .
\end{equation}
We will study the theory in the $d\rightarrow s$ sector.

The effective Lagrangian obtained from eq.~(\ref{slag}) after integrating
out the heavy scalar has the form
\begin{equation}\label{elag}
  {\cal L}_{eff}= {\lambda^2\over M^2}\,c_S(\mu)\ \bar s \tau\
  \bar e d + {\lambda^2\over  M^2}\,c_T(\mu)\ \bar s \sigma_{\mu\nu}
  \tau\ \bar e\ \sigma^{\mu\nu} d
  - {\lambda^2\over M^4}\, d_S(\mu)\ \bar s \tau\  D^2\left[\ \bar e d
  \ \right]+ \ldots
\end{equation}
where the ellipsis indicates additional operators of dimension 6 and higher.
The scale $\mu$ is the scale at which the effective Lagrangian is computed
from the full theory, and is usually chosen to be $\mu=M$.  The effective
Lagrangian at zeroth order in the strong interactions is computed by
equating the $d \tau\rightarrow s e$ scattering amplitude obtained from
eq.~(\ref{elag}) with that obtained by expanding the scalar exchange graph
in the full theory (Fig.~4) in a power series in $1/M^2$, to give $c_S=1$,
$c_T=0$, and $d_S=1$.

We now consider matching onto the effective theory at high orders in
$\alpha_s$, and concentrate on the coefficient of the operator $O_{T}\equiv
\bar s \sigma_{\mu\nu} \tau\ \bar e \sigma^{\mu\nu} d$.  This is a
particularly convenient operator because its coefficient $c_T$ is zero at
tree level, and at one loop in the full theory it only receives a
contribution from the graph in Fig.~5.  The other graphs do not have the
correct $\gamma$-structure to contribute to $c_T$, and contribute only to
the scalar amplitude.  $c_T$ is obtained by equating the tensor scattering
amplitudes in the full and effective theories, and its Borel transform is
computed using the techniques discussed in Section~2.

The Borel transform of the tensor scattering amplitude $A_T$ (at zero
momentum transfer) in the full theory is obtained by evaluating the Feynman
graph in Fig.~5,
\begin{equation}\label{fullgraph}
   B\left[A_T^{\rm full}\right]={16 \pi \lambda^2\over 3 N_f}
   \left({\mu^2\over e^C}\right)^u
   \bar U_s \sigma_{\mu\alpha} U_\tau\ \bar U_e \sigma_{\nu \beta} U_d\
   \int {d^4 k\over (2 \pi)^4}\ {k^\alpha k^\beta \left( k^\mu k^\nu - k^2
   g^{\mu \nu}\right)\over \left(k^2-m^2\right)^2
\left(k^2-M^2\right)\left(-k^2\right)^{2+u}}
\end{equation}
using the gluon propagator Eq.~(\ref{borelchain}).  The $U_i$ are Dirac
spinors for the external fermion lines, and the quarks have been given a
common mass $m$ to regulate the infrared behavior of the diagram.
Performing the $k$ integral gives
\begin{eqnarray}\label{fullgraphii}
   B\left[A_T^{\rm full}\right]&=&{i \lambda^2\over 12 \pi  N_f}
   \left({\mu^2\over e^C}\right)^u
   \bar U_s \sigma^{\mu\nu} U_\tau\ \bar U_e \sigma_{\mu\nu} U_d
   \ {\Gamma(u-1)\ \Gamma(2-u)}\\
   &&\qquad \left[ { \left(M^2\right)^{1-u} - \left(m^2\right)^{-u}
   \left[\left(1-u\right)M^2+ u m^2\right]\over \left(M^2-m^2\right)^2}
\right].\nonumber
\end{eqnarray}
Despite the appearance of $\Gamma (u-1)$ in eq.~(\ref{fullgraph}), the
amplitude is finite at $u=0$ and $1$ as the term in the square brackets
vanishes at these points, but it has singularities at $u=2, 3, \ldots$.
However, we stress that renormalon singularities in Eq.~(\ref{fullgraphii})
are of no interest, since we are not going to attempt to calculate
scattering amplitudes in perturbation theory.  Any attempt to do so will of
course face serious infrared problems.  We are only interested in using
perturbation theory to calculate the coefficient functions in
Eq.~(\ref{elag}), which requires us to subtract the corresponding amplitude
calculated in the effective theory.

To match onto the effective theory, we expand (\ref{fullgraphii}) in a
power series in $1/M$:
\begin{eqnarray}\label{fullcT}
   B[A_T^{\rm full}]  &=&{i \lambda^2\over 12 \pi  N_f} \left({\mu^2\over
   e^C}\right)^u \bar U_s \sigma^{\mu\nu} U_\tau\ \bar U_e \sigma_{\mu\nu} U_d
   \ {\Gamma(u-1)\ \Gamma(2-u)} \\
   &&\qquad \times \Biggl[{\left(M^2\right)^{-u}\over M^2} -
   \left(1-u\right){\left(m^2\right)^{-u}\over M^2}+\ldots\Biggr],\nonumber
\end{eqnarray}
where we have only retained terms up to order $1/M^2$, since $c_T/M^2$ is
the coefficient of a dimension six operator.  It is perhaps useful at
this point
to relate this to the standard perturbation series for $A_T^{\rm full}$,
\begin{equation}
A_T^{\rm full}=i{\lambda^2\over 12\pi N_f M^2}\ \bar U_s
\sigma^{\mu\nu}
U_\tau\ \bar U_e \sigma_{\mu\nu} U_d\
\sum_{n=0}^\infty a_n(\alpha_s N_f)^{n} + {\cal O}({1\over M^4}).
\end{equation}
Taking
the appropriate derivatives of (\ref{fullcT}) gives for the first few terms
of the series
\begin{eqnarray}
  a_0&=&0, \nonumber \\
  a_1&=&-1-2\log(m/M), \nonumber\\
  a_2&=&-{b_0}\left[C+2(1+C)\log(m/M)-2\log(\mu/M)\right. \nonumber \\
       &&\qquad\left.+2\log^2(m/M)- 4\log(m/M)\log(\mu/M)\right], \nonumber \\
  a_3&=& ...
\end{eqnarray}

The tensor scattering amplitude in the effective theory is computed from
the loop correction to the lowest order operator $c_S$,
\begin{equation}\label{eigraph}
   B\left[A_T^{\rm eff}\right]= -{16 \pi \lambda^2\over 3 N_f}
   \left({\mu^2\over e^C}\right)^u
   \bar U_s \sigma_{\mu\alpha} U_\tau \bar U_e \sigma_{\nu \beta} U_d\
   \int {d^4 k\over (2 \pi)^4}\ {k^\alpha k^\beta \left( k^\mu k^\nu - k^2
   g^{\mu \nu}\right)\over \left(k^2-m^2\right)^2 M^2\left(-k^2\right)^{2+u}},
\end{equation}
where we have used the tree-level value $c_S=1$ in evaluating the graph.
This result is the same as that obtained by setting the scalar propagator
in eq.~(\ref{fullgraph}) to $1/M^2$, since the $c_S$ term in the effective
Lagrangian reproduces this piece of the four-Fermi vertex. Evaluating the
$k$ integral gives
\begin{equation}\label{eigraphii}
   B[A_T^{\rm eff}] = {i \lambda^2\over 12 \pi  N_f} \left({\mu^2\over
   e^C}\right)^u \bar U_s \sigma^{\mu\nu} U_\tau\ \bar U_e \sigma_{\mu\nu} U_d
   \ {\Gamma(u)\ \Gamma(2-u)}\ {\left(m^2\right)^{-u}\over M^2}.
\end{equation}
This reproduces the $\left(m^2\right)^{-u}$ term in eq.~(\ref{fullcT}),
including the
entire $u$ dependence.
Comparing eq.~(\ref{fullcT})
with eq.~(\ref{eigraphii}), we obtain
\begin{equation}\label{match}
   B[c_T\left(\mu\right)]={\lambda^2\over 12 \pi  N_f} \left({\mu^2\over e^C
}\right)^u \
   {\Gamma(u-1)\ \Gamma(2-u)}\ {\left(M^2\right)^{-u} \over M^2}.
\end{equation}
Note that any dependence on the quark mass $m$ has dropped out of
(\ref{match}), so that the matching condition is independent of the
infrared regulator.  Therefore, in terms of the original perturbation
series
\begin{equation}
c_T(\mu)={\lambda^2\over 12\pi N_f}\ \sum_{n=0}^\infty c_n(\alpha_s
N_F)^{n},
\end{equation}
all large logarithms of $m/M$ have dropped out of the matching
conditions:\footnote{The logarithms of $m/M$ are reproduced in the effective
theory by scaling the operators from $\mu=M$ to low energies.}
\begin{eqnarray}
  c_{0}&=&0, \nonumber \\
  c_1&=&C^\prime-2\log(\mu/M), \nonumber \\
  c_2&=&-{b_0}\left[C^{\prime\prime} +2 C \log(\mu/M)-
    2 \log^2(\mu/M)\right], \nonumber \\
  c_3&=&...
\end{eqnarray}
(where $C^\prime$ and $C^{\prime\prime}$ are scheme-dependent
renormalization constants).  However, despite the fact that the individual
terms $c_i$ in the expansion of $c_T$ are now well-defined, the expression
(\ref{match}) has poles at $u=0, 1, 2, ...$.  The pole at $u=0$ is removed
by renormalization \cite{benekebraun}, but the renormalons at $u=1, 2, ...$
correspond to ambiguities of order $(\Lambda_{\rm QCD}/M)^{2u}$ in the
coefficient function $c_T(\mu)$.  Note that these are different from the
singularities in the expression (\ref{fullgraphii}).  The singularity at
$u=1$ in (\ref{match}) is not present in (\ref{fullgraphii}), while the
coefficients of the singularities at $u=2, ...$ also differ.  For the
singularity at $u=n$ we find
\begin{eqnarray}\label{singularities}
  &B&[A_T^{\rm full}]\sim{1\over M^2}\times
  {(-1)^n (n-1)\over u-n}\left({\mu^2\over m^2}\right)^n\ \sum_{i=1}^{n-1}{
  (n-i)\left(m^2\over M^2\right)^{i-1}}
  +...
  \nonumber \\
\noalign{\smallskip}
  &B&[c_T(\mu)]\sim -{(-1)^n\over u-n}\left({\mu^2\over M^2}\right)^n+ ...
\end{eqnarray}
where the ellipses denote terms regular at $u=n$.  Note that the singular
piece of $B[A_T^{\rm full}]$ at $u=n$ has no term proportional to
$(\mu^2/M^2)^n$.

We now consider using the effective Lagrangian in Eq.~(\ref{elag}) to
calculate the cross-section for the spin-flip scattering process $\tau
d\rightarrow e s$.  This is of course not a physical process, since the $d$
and $s$ quarks are not physical asymptotic states, and the perturbatively
calculated rate $A_T^{\rm full}$ exhibits serious infrared problems, as is
clear from Eq.~(\ref{fullgraphii}).  However, we may use this process to
demonstrate that physical predictions in the effective theory are
well-defined, by demonstrating that the renormalon ambiguities cancel
between the coefficient function $c_T(\mu)$ and the (perturbatively
computed) matrix elements of higher dimension operators.  Since the graphs
which cancel the ambiguity also occur for the matrix elements of physical
hadrons, this cancellation will also take place for physical amplitudes
such as $\tau\rightarrow K^0 e$.

The order $1/M^2$ contribution to $\tau d\rightarrow s e$ process from
the operator $O_T$ is
\begin{equation}\label{ampli}
   {\lambda^2\over 12 \pi  N_f} \left({\mu^2\over e^C }\right)^u \
   {\Gamma(u-1)\ \Gamma(2-u)}\ {\left(M^2\right)^{-u} \over M^2}\
   \bar U_s \sigma^{\mu\nu} U_\tau\ \bar U_e \sigma_{\mu\nu} U_d.
\end{equation}
Writing $u=1+\Delta u$, one finds that the singularity at $u=1$ is
\begin{equation}\label{amplising}
   {\lambda^2\over 12 \pi  N_f} \left({\mu^2\over e^C }\right) \
   {1\over M^4}\
   \bar U_s \sigma^{\mu\nu} U_\tau\ \bar U_e \sigma_{\mu\nu} U_d\
   {1\over \Delta u}
\end{equation}
which is of order $1/M^4$.  The renormalons at $u=2, \ldots$ produce
singularities of order $1/M^6, \ldots$.  Since we have only computed the
effective Lagrangian to order $1/M^4$, we can ignore the renormalon
singularities at $u\ge 2$, and the only singularity that is relevant to the
order we are working is the one at $u=1$.  This singularity is canceled by
a singularity in the Borel-transformed matrix element of the $d_S$
operator, which is also of order $1/M^4$.  The matrix element of the $\bar
s \tau\ D^2\left[\ \bar e d \ \right]$ operator between quark states is
evaluated using the graphs of Fig.~6, where only the first graph
contributes to the spin-flip scattering amplitude
\begin{equation}\label{eiigraph}
  -{16 \pi \lambda^2\over 3 N_f}
   \left({\mu^2\over e^C}\right)^u
   \bar U_s \sigma_{\mu\alpha} U_\tau\ \bar U_e \sigma_{\nu \beta} U_d\
   \int {d^4 k\over (2 \pi)^4}\ {k^\alpha k^\beta \left( k^\mu k^\nu - k^2
   g^{\mu \nu}\right)k^2\over \left(k^2-m^2\right)^2
M^4\left(-k^2\right)^{2+u}}.
\end{equation}
Evaluating the $k$ integral gives
\begin{equation}\label{eiigraphii}
   -{i \lambda^2\over 12 \pi  N_f} \left({\mu^2\over
   e^C}\right)^u \bar U_s \sigma^{\mu\nu} U_\tau\ \bar U_e \sigma_{\mu\nu} U_d
   \ {\Gamma(u-1)\ \Gamma(3-u)}\ {m^2\left(m^2\right)^{-u}\over M^4}.
\end{equation}
Expanding around $u=1$ gives the singular term
\begin{equation}\label{eiigraphsing}
   B[A_T^{(1)}] = -{i \lambda^2\over 12 \pi  N_f} \left({\mu^2\over
   e^C}\right) \bar U_s \sigma^{\mu\nu} U_\tau\ \bar U_e \sigma_{\mu\nu} U_d
   \ {1\over M^4}\ {1\over\Delta u}.
\end{equation}
This is precisely the negative of eq.~(\ref{amplising}), so that the
singularity cancels in the total amplitude, which is the sum of the two
terms eq.~(\ref{ampli}, \ref{eiigraphii}).

Clearly, at $u=2, 3, ...$ similar cancellations will take place with
$1/M^6, 1/M^8, ...$ operators.  This is simply because the singularities
in $B[c_T(\mu)]$ are not found in the scattering amplitudes in the full
theory, Eq.~(\ref{singularities}). Since the full and effective theories
are, by construction, identical up to the order to which the effective
theory has been defined, the singularities must cancel between matching
conditions and the matrix elements of higher dimension operators in the
effective theory, as they do at $u=1$.

\section{Conclusions}

Perturbatively calculated matching conditions in an effective field theory
suffer from renormalon ambiguities.  However, we have argued that any
ambiguity in a physical quantity is always higher order in $1/M$ than the
effective theory has been defined and is therefore of no consequence.  In
practice, one calculates matching conditions to a given number of loops,
and from physical measurements then determines the value of nonperturbative
matrix elements.  The cancellation of renormalon ambiguities in physical
observables then simply means that, although the values obtained for the
nonperturbative matrix elements will depend sensitively on the number of
loops at which the theories are matched, relations between physical
quantities will not. If the unphysical parameters are extracted from
observables at a given order in $\alpha_s$, then they can be used to
predict other observables to the same order in $\alpha_s$, as was done for
example for the extraction of $|V_{bc}|$ in \cite{ls,bu}. Inclusive and
exclusive semileptonic decays of hadrons containing a heavy quark
are free of renormalon ambiguities, regardless of the mass parameter of
the $1/m_0$ expansion. In addition, we have demonstrated renormalon
cancellation in an effective field theory other than HQET---the
four-Fermi theory.

\acknowledgements

We would like to thank A.~Falk for many useful discussions on this subject.
We would also like to thank H.~Georgi, I.~Rothstein, N.~Uraltsev,
A.~Vainshtein and M.~Wise for useful conversations.  A recent preprint by
Neubert and Sachrajda \cite{neubsach} draws similar conclusions about the
cancellation of renormalon ambiguities in HQET to those presented in this
paper.  Some of this work was carried out at the Weak Interactions Workshop
at the Institute for Theoretical Physics in Santa Barbara.  ML and M.J.S.
are grateful to the staff of the ITP and the organizers of the workshop.
MJS thanks the High Energy Physics groups at UC San Diego and Caltech, ML
thanks the High Energy Physics groups at UC San Diego and AM thanks the
Aspen Center for Physics for kind hospitality during the course of this
work.  This work was supported in part by the Department of Energy under
grant DOE-FG03-90ER40546 and by the National Science Foundation under grant
PHY89-04035.  A.M.  was also supported by PYI award PHY-8958081 from the
National Science Foundation.

\vfill\break\eject

\centerline{\bf Figure Captions}

\begin{enumerate}

\item[1] The bubble chain diagrams which are the leading contribution
to the renormalon as the number of light colored fermions is large,
$N_f\rightarrow \infty$.

\item[2] The one-loop contribution to the matching of a higher dimension
operator.  The coefficient of the operator in the effective Lagrangian is
$c_0+c_1\alpha_s+c_2\alpha_s^2+...$ .   This series will have IR renormalons
due to the incomplete cancellation of the soft gluons in the two graphs.

\item[3]  The loop graph with the Borel transformed
gluon propagator contributing to the $v^\mu$ form factor in
$\Lambda_b\rightarrow\Lambda_c e\nu$.

\item[4] The tree-level exchange of a heavy colored scalar.
It only contributes to $c_S(\mu)$ and $d_S(\mu)$ and not to
$c_T(\mu)$.

\item[5] The leading contribution to $c_T(\mu)$ arises at one
loop in QCD. The blob on the gluon propagator denotes the bubble sum of
light quark loops. The external quarks have been given a small mass $m$
which  regulates the IR behavior of the graph.

\item[6] The contribution to the scattering amplitude from
the $(iD)^2$ operator which is ${\cal O}(1/M^4)$. Only the first diagram
contributes to the spin-flip amplitude at order $1/N_f$. The ambiguity at
$u=1$ in this matrix element cancels that in $c_T(\mu)$.

\end{enumerate}
\vfill\eject
\insertfigsmall{}{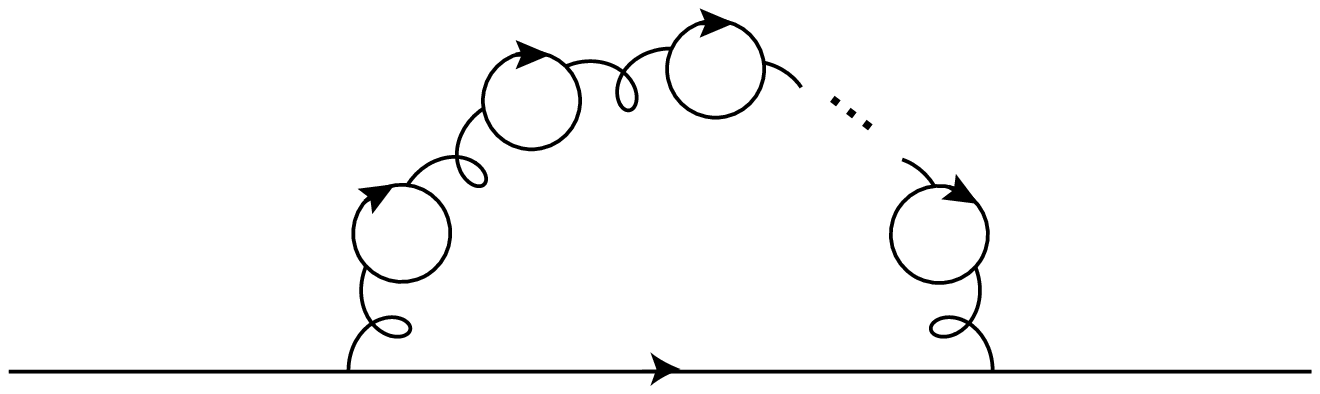}
\centerline{Figure 1.}
\vskip 0.75in
\insertfig{}{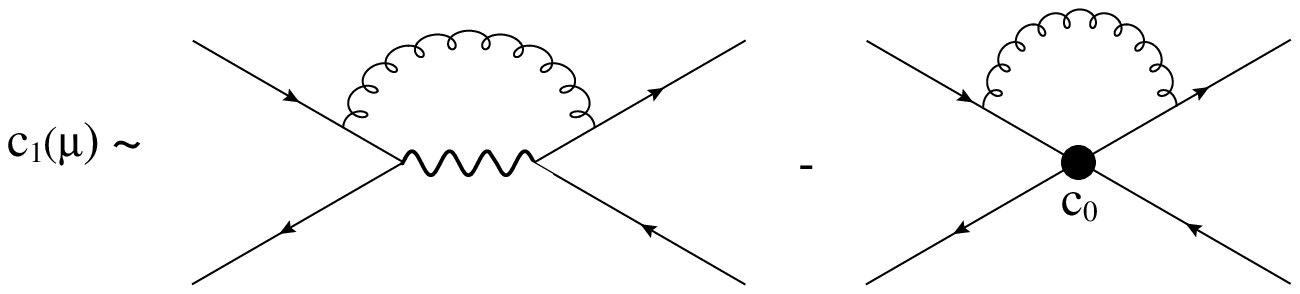}
\vskip -0.5in
\centerline{Figure 2.}
\vskip 0.75in
\insertfig{}{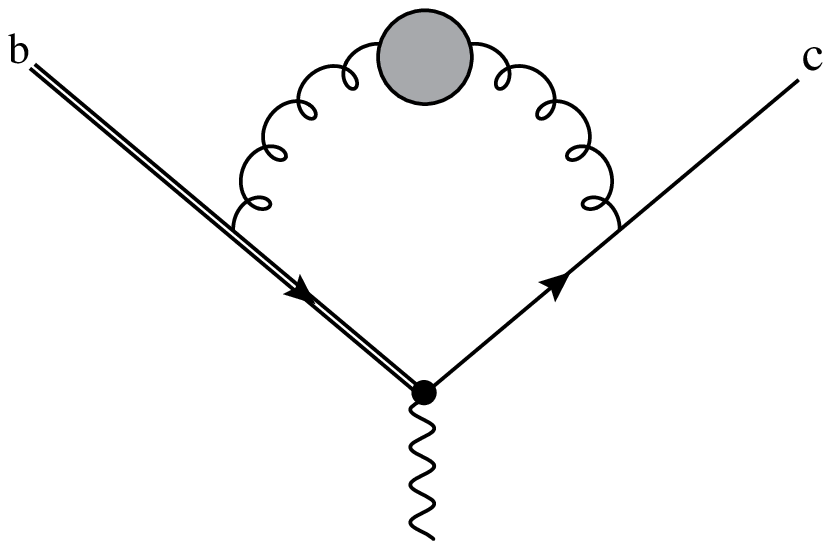}
\centerline{Figure 3.}
\vskip .5in
\insertfig{}{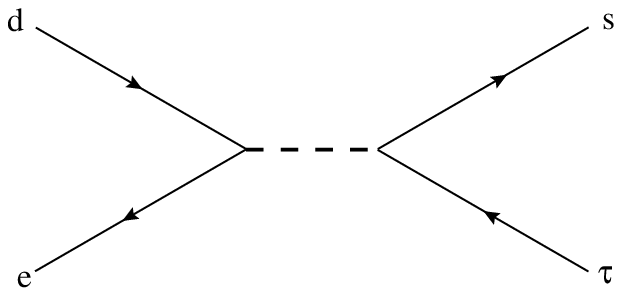}
\centerline{Figure 4.}
\vskip 1in
\insertfig{}{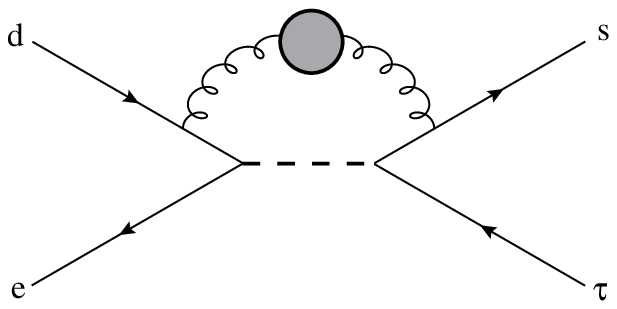}
\centerline{Figure 5.}
\vskip .5in
\insertfig{}{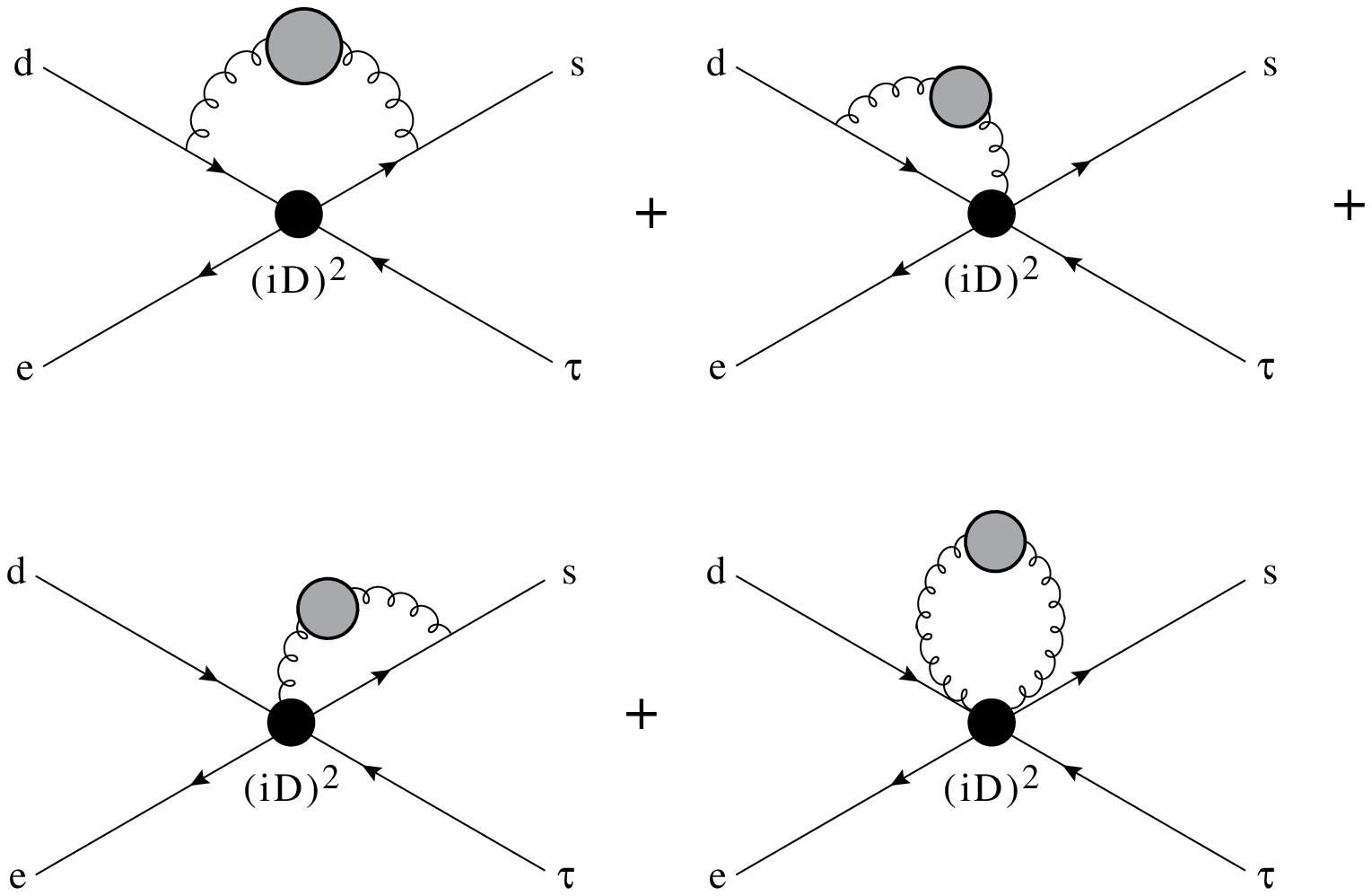}
\centerline{Figure 6.}

\end{document}